\documentclass[a4paper,11pt]{article}
\usepackage{jcappub} 
\usepackage{url} 
\usepackage{multirow}
\usepackage{xcolor,colortbl}

\usepackage{amsmath,amssymb} 
\usepackage{mathrsfs} 
\usepackage{lineno}
\usepackage{amsmath}

\title{\boldmath  Shedding Light on Mirror Dark Matter Using CDEX}

\author[a,b]{ M. Tousif Raza}

\affiliation[a]{Department of Physics, Oklahoma State University, Stillwater, 74078, OK, USA\\
}
\affiliation[b]{Department of Physics \& Astronomy, University of New Mexico, Albuquerque, 87131, NM, USA\\
}

\emailAdd{tousif@okstate.edu}


\abstract{Mirror dark matter, a hidden-sector counterpart to the Standard Model with identical particle content, could be detectable through photon-mirror-photon kinetic mixing $(\epsilon)$. We present new constraints from CDEX-10 low-energy recoil data, establishing $\epsilon < 2 \times 10^{-10}$ for mirror nuclear dark matter in the 10–56 GeV mass range. Projections for CDEX-300 (with 95 days of exposure) show significantly improved sensitivity, reaching $\epsilon < 3.75 \times 10^{-11}$ for a local mirror halo temperature below 0.3 keV, surpassing existing constraints from LUX and positronium decays. These results demonstrate CDEX-300's unique capability to probe kinetic-mixing parameter space, providing stringent constraints on mirror dark matter models.}

\begin{document}
\maketitle
\flushbottom
\section{Introduction}
  Observations of galactic rotation curves and the Cosmic Microwave Background (CMB) have provided strong evidence for the existence of a new form of matter in the universe, popularly known as dark matter (DM)~\cite{Rubin:1978kmz,Planck:2018vyg}. DM non-baryonic and cannot be explained within the framework of the Standard Model (SM). The nature of DM is unknown to date and is considered one of the most important questions to the physics community.  In the past, considerable efforts have been made to uncover the properties DM and its cosmological origins. Currently, a vast experimental program is underway, with several next-generation experiments under construction to advance the search for DM~\cite{XENON:2023cxc, PhysRevLett.131.041002, PandaX-4T:2021bab,DARWIN:2016hyl,Aalseth_2018}. Traditionally DM direct detection experiments operate deep underground to ensure the maximum background rejection and the invaluable channel for discovery is the nuclear and electronic recoil events induced by DM particles in the detector. Many of the mid and large-scale experiments are searching for weakly interacting massive particles (WIMPs), a theoretically well-motivated candidate but a significant range of WIMP parameter space has already been excluded by many of the direct detection experiments~\cite{Roszkowski:2017nbc, LZ:2022lsv}. As a result, interest in other theoretical models has increased. Optimistically there is a plethora of hypotheses on dark matter \cite{Bertone:2004pz, Bergstrom:2009ib}. One such hypothesis is based on a mirror-dark sector with mirror symmetry. For the first time, the concept of mirror symmetry was introduced in understanding the parity violation in weak interaction 
of SM~\cite{Lee:1956qn}.  It has been emphasized that if mirror symmetry is respected by nature, then there is likely to exist a mirror world that would completely replicate the ordinary world. In this scenario, one would expect mirror electrons and mirror quarks to be the fundamental constituents of mirror matter, and mirror atoms are made of mirror nucleon (i.e. mirror proton, mirror neutron). Similar arguments for the existence of a mirror dark sector come from
the study of neutron oscillations see Ref.~\cite{Berezhiani:2011da}. Mirror particles could be the potential candidates for DM and this can be restored  or ruled out by the direct detection experiments. In this work, it is assumed that mirror matter forms part of dark matter in the galaxy with a number density lower than the standard cold collisionless DM. We address physics related to the mirror electromagnetic and standard electromagnetic interaction with a possible massless photon-mirror-photon kinetic mixing hence any
features of mirror weak and strong along with standard weak and strong interactions are irrelevant in our analysis. We extensively study the
impact of elastic mirror DM scattering on germanium target for a hypothetical process $(A^{\prime}+A\rightarrow A^{\prime}+A)$ relevant for CDEX-10, where CDEX stands for China Dark Matter Experiment operating in China Jinping Underground Laboratory. To the best of our knowledge of late only a few studied have been conducted in search for mirror DM in direct detection therefore our present study shed light on the physics of the mirror dark sector utilizing the current state-of-the-art experiment CDEX-10 and an forthcoming CDEX-300. 

We organized our work as follows. In section $\S$~\ref{sec:Mirror dark matter in the galaxy} we begin with an overview of mirror dark matter in the galaxy. In section $\S$~\ref{sec:Halo model}, we revisit the Halo model and associated astrophysical and cosmological parameters relevant to mirror DM analysis, in addition, we discuss an Empirical model of DM velocity distribution in the mirror sector. In section $\S$~\ref{sec:Scattering formalism}, we thoroughly discuss the scattering formalism of mirror dark matter with germanium where we explore mirror nuclear form factors and other ingredients that appear in the scattering rates. 
In section $\S$~\ref{sec:Direct detection} We bridge theoretical predictions with experimental details and present a statistical analysis. We summarize our results in section $\S$~\ref{sec:Results}. Finally, conclude in Section $\S$~\ref{sec:Conclusion}. 

\section{Mirror dark matter in the galaxy}
\label{sec:Mirror dark matter in the galaxy}
It is believed that the mirror DM is isomorphic to the Standard
Model (SM) that lives in the hidden dark sector and can participate in kinetic mixing interaction with the SM. The relevant Lagrangian density for the dark sector and the kinetic mixing between SM and dark sector can be described as~\cite{Foot:2016wvj, Berezhiani:2000gw, Carlson:1987si},

\begin{equation}
\mathcal{L_T}=\mathcal{L_{SM}}(e,\mu,d,
\gamma,\cdots)+\mathcal{L_{D}}(e',\mu',d',
\gamma',\dots)+\mathcal{L_{M}}
\end{equation}
\begin{equation}
    \mathcal{L_{M}}=\frac{\epsilon}{2}F^{\mu\nu}F'_{\mu\nu}
    \label{eq:2.2}
\end{equation}
where $\mathcal{L_{M}}$ represents the massless standard photon and massless mirror photon kinetic mixing.
Here $F^{\mu\nu}=\partial^\mu A^\nu-\partial^\nu A^\mu$ is the standard electromagnetism field tensor  and $F'_{\mu\nu}=\partial^\mu A'^\nu-\partial^\nu A'^\mu$ is the mirror electromagnetism field tensor respectively. $\mathcal{L}_{SM}$ has a gauge symmetry SU(3)$_C\otimes$ SU(2)$_L\otimes$U(1)${_Y}$ and
the mirror dark sector represented by the lagrangian $\mathcal{L_D}$ is presumed to have mirror particles  with a gauge symmetry SU(3)$^{\prime}_C\otimes$ SU(2)$^{\prime}_L\otimes$U(1)$^{\prime}{_Y}$. In the case of the mirror sector with a new gauge field U(1)$^{\prime}{_Y}$, it is believed mirror particles can interact with the particles in standard sector U(1)$_Y$ via kinetic mixing  $\mathcal{L_{M}}$ as described in (\hspace{-0.15cm}~\ref{eq:2.2}) and induce induces a mirror charge $\epsilon$Z$'$e$'$ that can be measured in the direct detection experiments ~\cite{Holdom:1985ag}. Here our primary focus is on the single elemental  mirror DM
 namely H$^{\prime}$, He$^{\prime}$, C$^{\prime}$, O$^{\prime}$, Fe$^{\prime}$, etc., therefore mirror dark matter in compound form and their existence in the galaxy is beyond the scope of this work. We highlight the possible benchmark values of mirror dark matter compositions in our solar neighborhood ~\cite{Cerulli:2017jzz}.
 \begin{table}[h!]
\centering
\begin{tabular}{ |c|c|c|c|c|c|} 
\hline\hline Mirror Matter & H$'(\%)$ & He$'(\%)$& C$'(\%)$ &O$'(\%)$ & Fe$'(\%)$ \\
\hline
{H$'$, He$'$} &25 & 75&-&-&- \\ {H$'$, He$'$, C$'$, O$'$}
& 12.5 & 75 &7&5.5&-\\ {H$'$, He$'$, C$'$, O$'$, Fe$'$}
& 20 & 74&0.9&5&0.1\\ 
\hline\hline
\end{tabular}
\caption{Mirror matter abundance in the solar neighborhood.}
\label{tab: Tab.1}
\end{table}

Mirror matter heavier than primordial  H$'$, He$'$ can be present in the halos around spiral galaxies under several assumptions. One such assumption is motivated by the fact that heavy mirror atoms (i.e. Fe$'$)  are produced in stars and 
propelled in the galaxy by supernova explosions thereby validating the idea of having heavy mirror atoms in the halos. The heavier mirror matter namely C$'$, O$'$, and  Fe$'$  are necessarily required for stability of halo \cite{Foot:2013uxa}. The idea of mirror electrons is motivated by the fact that mirror
photons will heat the mirror DM particles to make them ionized as a result dark plasma of mirror electrons and mirror nuclei i.e. H$'$, He$'$, O$'$, Fe$'$, etc. is  achieved~\cite{raffelt1996stars, Foot:2014uba,2017PhLB..766...29C, Foot:2014mia}. Since Mirror dark plasma is a collection of mirror negative and mirror positive charges eventually these charged states are dissipative and expected to evolve in time maintaining a steady state on balancing heating and cooling rates. Assuming mirror dark plasma is dominated by a single elemental composition mainly mirror helium in the Milky Way halo, the parameter space of the kinetic mixing can be inferred as~\cite{Berezhiani:2003wj, Foot:2016wvj};

\begin{equation}
     10^{-11}\leq \epsilon \sqrt{\xi_{A'}}\leq  10^{-9}
     \label{eq: 2.3}
\end{equation}
 Here $\xi_{A'}$ is the mass fraction of mirror matter that can be read from Table~\ref{tab: Tab.1} and it is related to the mirror number density.
 We thoroughly examine this kinetic mixing bound from Eq.(\hspace{-0.1cm}~\ref{eq: 2.3}) employing low-energy recoil data from CDEX-10 and we perform simulation for forthcoming CDEX-300 to infer the ultimate capability of this experiment in studying mirror DM.




\section{Halo model}
\label{sec:Halo model}
In a canonical isothermal halo, mirror DM particles are gravitationally bound, and their velocity distribution can be approximated as collisionless and Maxwellian. We assume mirror dark-matter particles are in thermal equilibrium with temperature $T^\prime$ and velocity dispersion $v_0$, where the thermal equilibrium state is a direct measure of temperature in halo. While the collisionless treatment is an idealization, its validity depends on the mean free path of mirror DM particles typically a kiloparsec ($\sim$ kpc), which effectively justifies the non-collisional approximation. Under such assumption, the mirror DM velocity distribution takes the following form~\cite{McCabe:2010zh},

\begin{equation}
f(v,v_E) =\begin{cases}
          \frac{1}{N} \left(e^{-{(v+v_E)^2}/{v_0^2}}\right) & v <v_{c} \\
          0  & v>v_{c}
          \end{cases}
\end{equation}
where $v=v_{min}$ is the minimum speed and $v_c$ is the cut-off speed of the mirror DM particles similar to the galactic escape speed $(v_{esc})$~\cite{Chua:2013mc}. $N$ is the normalization constant can be found by setting $\int f(v)d^3v=1$ and ${v}{_E}$(t) represents the Earth velocity with respect to the
galactic rest frame while time dependence carries the meaning of Earth's revolution around the Sun~\cite{Bramante:2016rdh}.
Including the effect of $v_E(t)$ mirror DM velocity integral yields,

\begin{equation}\zeta_{A'}\left(v_{min}(E_R),t\right)=\int_{v_{min}}^{\infty} f( v, v_E) \frac{d^3v}{v}
\label{eq: 3.2}
\end{equation}

\begin{equation}v_E(t)=v_s+bv_\oplus cos[w(t-t_0)]
\end{equation}
 provided $v_s$ is the velocity of the Sun in the galactic rest frame, i.e. $v_s\approx 232$ km/s. We use the speed of the Earth's orbit to be $v_\oplus=30$ 
 km/s ~\cite{Lee:2013xxa} and $\omega=\frac{2\pi}{T}$ with T$=1$ year, t$_0=152$ days and t is the experiment run-time.
 A simplified result of the velocity integral in (\hspace{-0.15cm}~\ref{eq: 3.2}) can be obtained in the limit $(v_c \rightarrow \infty)$ as,

\begin{equation}
\zeta_{A'}\left(v_{min}(E_R),t\right)=\frac{N}{2\pi^{3/2}v_0(A')^3v_E}\left[ erf(y_{m}+y_E)-erf(y_{m}-y_E)\right]
\end{equation}
with $y_m=\frac{v_{min}}{v_0(A')}$, $y_E=\frac{v_{E}}{v_0(A')}$ and normalization $N=\pi^{3/2} v_0(A')^3$. Here $v_0(A')$ is the velocity dispersion that can be determined from the hydrostatic equilibrium in the halo with local temperature $T^\prime=\frac{1}{2}{\bar m v_{rot}^2}$ given their rotational velocity velocity of the Milky Way $v_{rot}$. In principle, $v_{rot}$ may range  $190\leq v_{rot}\leq 250$ km/s \cite{2017PhLB..766...29C}.  In a completely ionized halo, mirror velocity dispersion reads,
\begin{equation}
v_0(A')=\sqrt{\frac{2T^\prime}{M_{A'}}}=\sqrt{\frac{\bar{m}}{M_{A'}}}v_{rot}
\label{eq:3.5}
\end{equation}

given that $A^\prime=e^\prime, H^\prime, He^\prime,\dots$ etc. In addition to the standard Maxwellian velocity distribution function (VDF), we utilize the empirical model of the DM velocity distribution\cite{Mao:2013nda},

\begin{equation}
f(\vec v) =\begin{cases}
           e^{-|v|/{v_0}}\left(v_{esc}^2-|v|^2\right)^{p} & |v|<v_{esc} \\
          0  & |v|>v_{esc}
          \end{cases}
\label{eq: 3.6}
\end{equation}

The function $f(\vec v)$ is composed of two terms: An exponential term encoding the halo's velocity anisotropy, and a cutoff term limiting particles bound to the Galactic potential. The circular speed $v_0$ is given by (\hspace{-0.15cm}~\ref{eq:3.5})  while the power-law index p (ranging 1–4) depends on the ratio $v_{rms}/v_{esc}$ as discussed in Ref.  \cite{Mao:2013nda}. Adopting $p=1.5$ we solve Eq. (\hspace{-0.1cm}~\ref{eq: 3.2}) and Eq.(\hspace{-0.1cm}~\ref{eq: 3.6}) with the normalization condition:
\begin{equation}
    \int 4\pi v^2 e^{\left(-{v}/{v_0}\right)} \left(v_{\text{esc}}^2 - v^2\right)^p \, dv = N
\end{equation}
Fig.~\ref{fig:Fig.1}(right panel) compares the resulting velocity integrals. While this empirical VDF is standard for cold dark matter studies, we adapt it to the mirror sector.  
In a homogeneous plasma consisting of H$'$,He$'$,e$'$, mean mass of mirror DM particles can be written as~\cite{Ciarcelluti:2010dm},
\begin{equation}
    \bar{m}=\frac{\sum n_{A'}M_{A'}}{\sum n_{A'}}=\frac{n_{He'}M_{He'}+n_{H'}M_{H'}+n_{e'}M_{e'}}{n_{He'}+n_{H'}+n_{e'}}
\end{equation} 
Assuming charge neutrality and He$^\prime$ dominated Milky Way mirror plasma, mirror electron number density $(n_{e'})$ satisfies the relation, 
\begin{equation}
n_{e'}=n_{H'}+2n_{He'}
\end{equation}
For the fully ionized case, the mirror electron number density in the plasma can be determined by,
\begin{equation}
    n_{e'}=\frac{\rho}{M_u}\sum_{Z=H',He'}\frac{X_Zn_Z}{A_Z}=\frac{\rho}{M_u}\left(1-\frac{Y_{He'}}{2}\right)
\end{equation}
given that $X_H+Y_{He'}\equiv 1$ where 
$n_Z$ denotes the number of free electrons contributed by atomic species Z (with 
$X_Z$ being their respective mass fractions in the mirror plasma) and $M_u$ is defined as 1 atomic mass unit (approximately equal to the hydrogen mass $m_{H^\prime}$). The mirror hydrogen ($m_{H^\prime}$) and helium ($m_{He^\prime}$) masses dominate over the mirror electron mass $(m_{e^\prime})$ satisfying
$M_{H},M_{He'}\gg 
M_{e'}$ and $M_{He'}=4M_{H'}$ applying the charge neutrality condition, the mean mass of mirror dark matter is given by,
\begin{equation}
    \bar m\approx \frac{M_H'}{2-\frac{5}{4}Y_{He'}}
\end{equation}
where $Y_{He'}$ is the mirror helium fraction~\cite{Ciarcelluti:2010dm}, \begin{equation}
Y_{He'}=\frac{n_{He'}M_{He'}}{n_{He'}M_{He'}+n_{H'}M_{H'}}
\end{equation}
Assuming the kinetic mixing $\epsilon\sim 10^{-9}$ and mirror helium fraction $Y_{He}'\equiv 0.74$, we find the mean mass of mirror DM particles to be $\bar{m}\approx 0.9$ GeV in a Milky Way halo for which local temperature below 1 keV is well expected. Therefore using the mean mass of $\bar m\approx 1$ GeV, we find that the lightest mirror DM particles attain higher velocities than their heavier counterparts. Also notice that  velocity dispersion is smaller
 in mirror halos than the distribution expected for collisionless cold dark matter (CDM). Usually, collisionless CDM i.e. WIMPs possess much higher velocity dispersion that approaches the galactic rotational velocity i.e.  $v_{0}\approx v_{rot}$. 

 In the galactic dark plasma, mirror electrons attain the highest velocity dispersion than mirror nuclei due to the fact that mirror electrons are significantly lighter than mirror nuclei. This ensures velocity integral for a mirror electron will be dominated by the contribution from mirror electrons itself. We may safely ignore Earth's velocity as it is negligible compared to typical velocity dispersion of mirror electron in halo. Therefore we write,
\begin{equation}
    \zeta_{e'}\left(v_{min}(E_R),t\right)=\int_{v_{min}}^{\infty} f_{e'}(v)\frac{d^3v}{v}
\end{equation}
provided velocity dispersion $v_0(e')=\sqrt{\frac{2T^\prime}{m_{e'}}}$ with  $N=\pi^{3/2} v_0(e')^3$ and $v_{min}$ is the minimum velocity of mirror electron available in the elastic scattering with the standard model electrons. Both mirror DM and mirror electron velocity integrals depend on
the astrophysical parameters with uncertainties that can importantly affect signal events. Besides interaction rates of mirror DM particles with the target nucleus have direct consequences to nuclear form factors $F^2(q)$, which contains
the nuclear physics properties i.e. effective radius and nuclear skin thickness of mirror DM nuclei as well as target nuclei. In the following section, we review mirror DM scatterings and briefly discuss all these aspects in quantifying the differential recoil rates.
\section{Scattering formalism }
We analyze mirror DM scattering with Standard Model targets in the non-relativistic regime, focusing on elastic scattering processes:  mirror nuclei scattering off target nuclei, and mirror electrons scattering off loosely bound atomic electrons. Within this framework, we can safely neglect scattering between target nuclei and mirror electrons.  
\label{sec:Scattering formalism}
\subsection{Mirror nuclear DM scattering rate}
\label{Mirror dark matter scattering rate}
If an incoming mirror nucleus with a minimum velocity $(v_{min})$ scatters off an ordinary nucleus  at rest given by the process,
\begin{equation}
    {N}^{\prime}+{N_T}\rightarrow {N}^{\prime}+{N_T}
    \label{eq:4.1}
\end{equation} 
   The momentum transfer $q\le 2\mu_Tv_{min}$ resulting in a nuclear recoil energy $E_R$. We ensure full kinematic coherence across the entire accessible range of momentum transfer by imposing the condition,
 
\begin{equation}
    \mu_T\ll \frac{1}{2v_{min}R}
\end{equation}

\begin{equation}
\mu_T= \frac{M_{A^{\prime}}M_T}{M_{A^{\prime}}+M_T}
\end{equation}
Where $\mu_T$ is the reduced mass of the two-body system i.e. mirror nuclei $N'$ and target $N_{T}$ and R denotes the effective nuclear radius of the target nucleus with mass $M_T$. For a mirror DM having mass $M_{A^{\prime}}$, a minimum incoming speed in laboratory frame can be found as,
\begin{equation}
    v_{min}=\sqrt{\frac{\left(M_T+M_{A^{\prime}}\right)^2E_R}{2M_TM_{A^{\prime}}^2}}
\end{equation}

\begin{figure}[htbp]
\centering
\includegraphics[width=.47\textwidth]{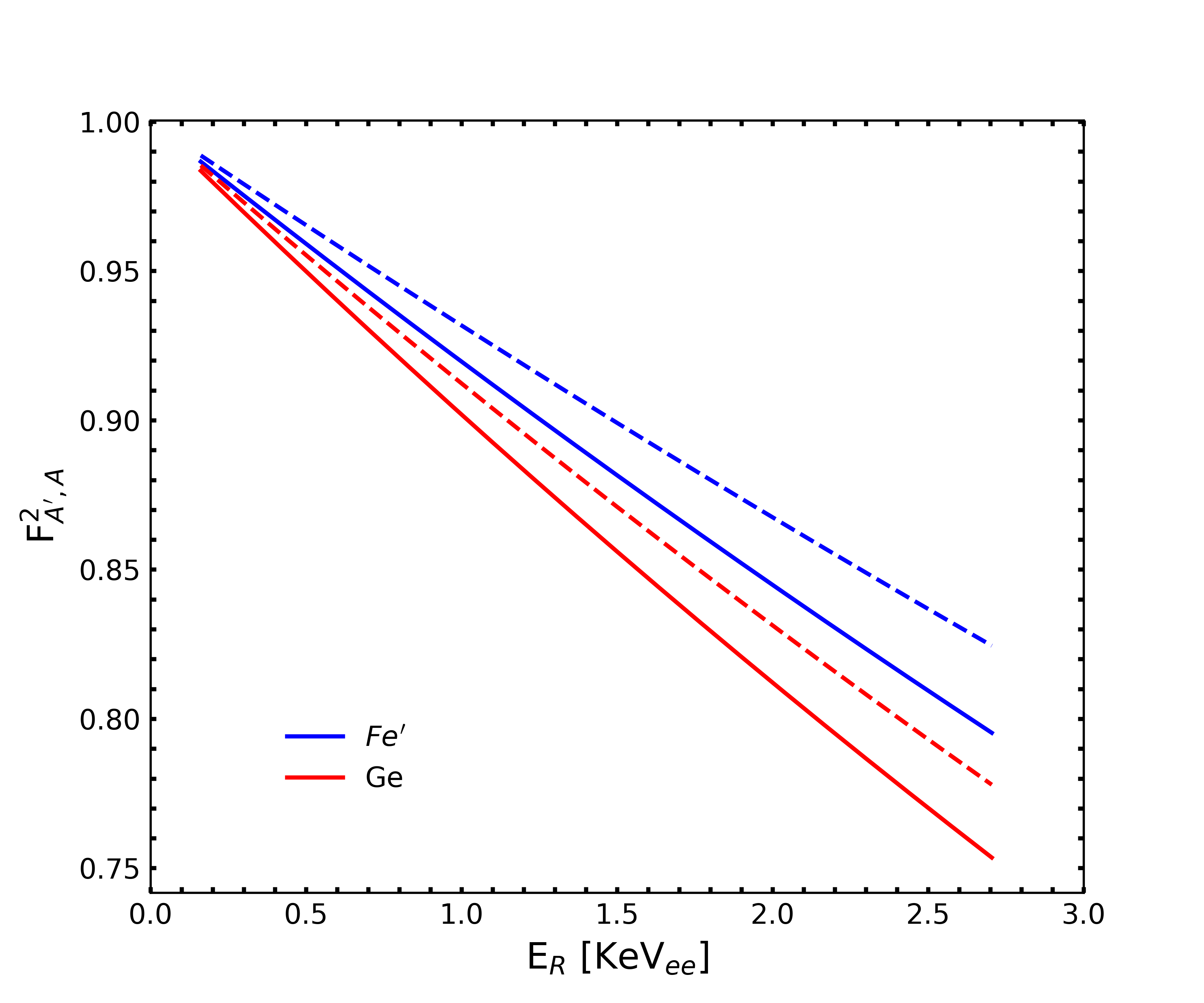}
\qquad
\includegraphics[width=.47\textwidth]{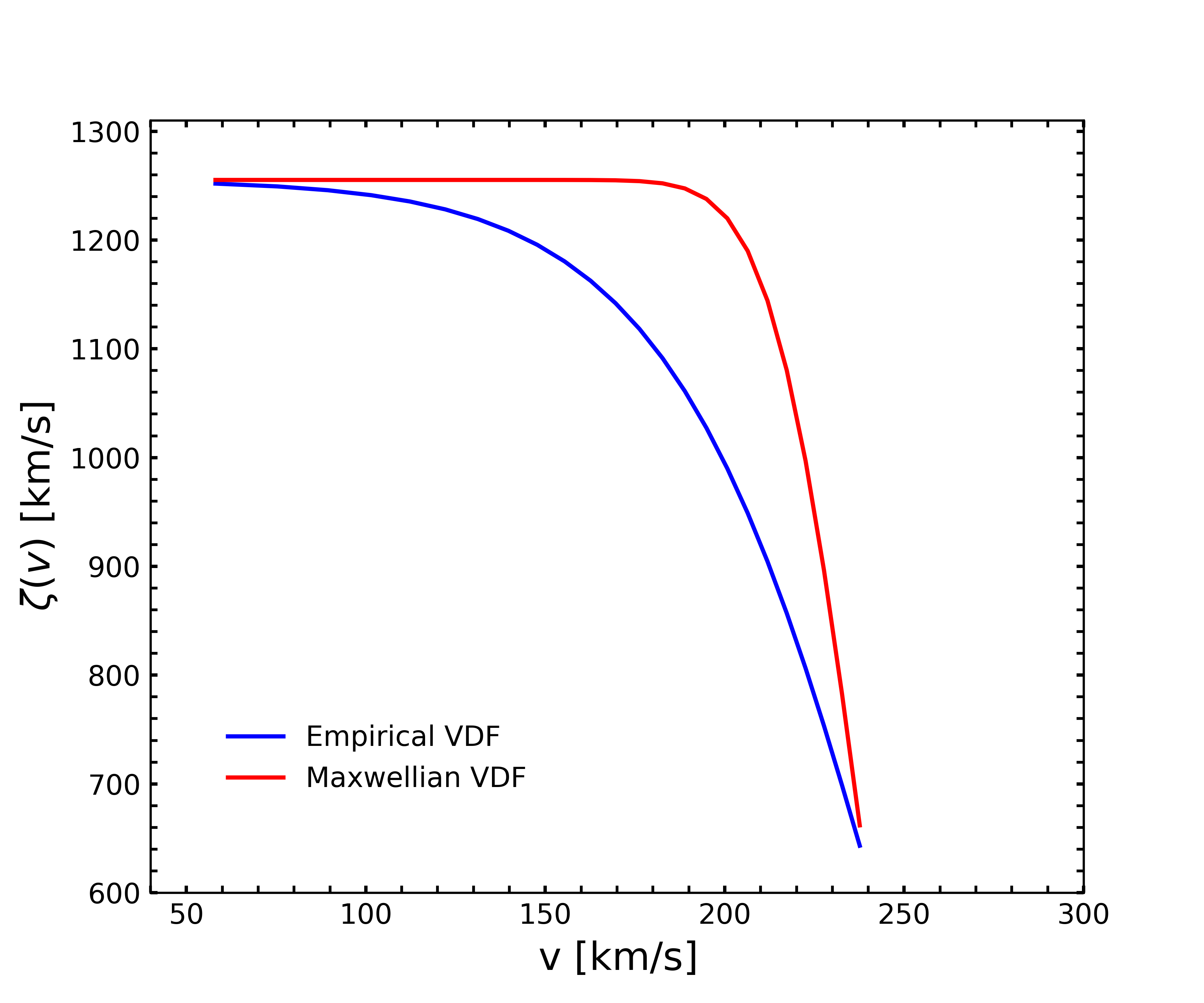}
\caption{Left: Form factors for mirror dark matter as a function of electron-equivalent recoil energy ($E_{ee}$). Right: Velocity distributions of mirror dark matter versus velocity. Both panels use the CDEX-10 threshold energy $E_{th} = 160$ eV$_{ee}$. Blue curves show mirror-sector form factors (solid: Helm/Lewin-Smith; dashed: Helm/Engel parameterizations), while red curves display the corresponding germanium nuclear form factors. \label{fig:Fig.1}}
\end{figure}

The scattering process described above is of Rutherford type, where a mirror nucleus with electric charge $Z^{\prime}$ interacts with an ordinary nucleus of charge $Z$ therefore the differential scattering cross-section can be estimated as follows~\cite{Foot:2013msa,Foot:2002xz}:
\begin{equation}\frac{d\sigma_{A^{\prime}}}{dE_R}=\frac{2\pi\epsilon^2\alpha^2 Z^2Z^{\prime^2} F_{A}^2F_{A^{\prime}}^2}{M_T E_R^2v_{min}^2}
\end{equation}
 Here, $\alpha$ denotes the fine-structure constant, while 
$F_A^\prime$ and
$F_A$ represent the form factors of the mirror and ordinary nuclei respectively. The scattering between target nuclei and mirror dark matter cannot be treated as point-like, as both possess extended nuclear charge distributions due to their finite sizes. To properly model this interaction, it is necessary to introduce a charge density distribution that captures the essential nuclear features and yields physically acceptable form factors. For the scattering process described in Eq. (\hspace{-0.15cm}~\ref{eq:4.1}), the form factors of both mirror and ordinary nuclei can be approximated by the spin-independent Helm form which is derived from the Fourier transform of a Woods-Saxon nuclear charge density distribution, characterized by R (nuclear radius) and s (surface thickness). The resulting form factor is a function of the momentum transfer q can be written as~\cite{ORRIGO2016414,PhysRev.104.1466},
\begin{equation}
F(q^2)=\frac{3j_1(qR)}{qR}e^{-\frac{(qs)^2}{2}}
\end{equation}
where $j_1$ is the spherical Bessel function of the first kind given by,
\begin{equation}
j_1(qR)=\frac{\sin(qR)-qR\,\cos(qR)}{(qR)^2}
\end{equation}
A choice of  R and s can be made as discussed by Engel~\cite{Engel:1992bf,Engel:1991wq},
\begin{equation}R=\sqrt{\tilde{R}^2-5s^2}
\end{equation}
with $\tilde{R}=1.2A^{1/3}$ fm and $s=1$ fm.
Alternatively, the parameters 
R and 
s can be determined using the Lewin-Smith parametrization~\cite{1996APh.....6...87L},
\begin{equation}
R=\sqrt{c^2+\frac{7}{3}\pi^2a^2-5s^2}
\end{equation}
 $s=0.9$ fm, $a=0.52$ fm and $c=(1.23\, A^{1/3}-0.6)$ fm.  The effective nuclear radius and skin thickness of mirror (or ordinary) nuclei are crucial parameters in computing the Helm form factors, as they directly scale with the differential scattering cross-section. For germanium nuclei, we observe variations in form factors as shown in left-panel (see Fig.\ref{fig:Fig.1}), though these lead to nearly identical scattering rates. In our analysis, we adopt the widely used Lewin-Smith parameterization.

 The recoil rate of Mirror DM particle off a SM target nuclei is given by, \begin{equation}\frac{dR_{A^{\prime}}}{dE_R}={N_T} n_{A^{\prime}}\int_{v_{min}} \frac{d\sigma_{A^{\prime}}}{dE_R}f(v,v_E)v d^{3}v
 \end{equation}
 
    \begin{equation}\frac{dR_{A^{\prime}}}{dE_R}={N_T}n_{A^{\prime}}\left(\frac{2\pi\epsilon^2\alpha^2 Z^2Z^{\prime^2} F_{A}^2F_{A^{\prime}}^2}{M_TE_R^2}\right)\zeta_{A'}\left(v_{min}(E_R),t\right)
    \label{eq:4.10}
 \end{equation}
We are primarily interested in estimating signal events caused by the maximum and minimum variations in Earth's velocity during the exposure run, which could potentially affect both the expected events and the kinetic mixing sensitivity. While the daily modulation effect is not investigated in our analysis, this phenomenon has been extensively studied in the context of DAMA's signal interpretation~\cite{DAMA:2008jlt, DAMA:2010gpn, Foot:2018xnx}. We denote the mirror dark matter density as,
\begin{equation}
n_{A^{\prime}}=\frac{\rho_{A^{\prime}}\xi_{A^\prime}}{M_{A^\prime}}
\end{equation}
The number density of mirror DM is necessarily lower than that of standard cold dark matter, scaled by the mirror sector's metal mass fraction in the galactic halo.
Following standard CDM conventions, we adopt a local mirror DM density $\rho_{A^{\prime}}=(0.3-0.4)$ GeV/cm$^{3}$ throughout this analysis ~\cite{PhysRevD.98.030001}.

\subsection{Mirror electron scattering rate}
\label{Mirror electron scattering}
Analogous to the mirror nuclear DM scattering, we study mirror electron scattering. These mirror electrons possess sufficient kinetic energy to escape the galactic plasma and undergo elastic scattering with loosely bound atomic electrons in the detector.
\begin{equation}
    {e}^{\prime}+e\rightarrow {e}^{\prime}+e
    \label{eq: 4.11}
\end{equation} 
The interaction constitutes a spin-independent Coulomb scattering involves momentum transfer between mirror electrons and loosely bound atomic electrons, generating keV-scale recoils detectable in direct detection experiments. By treating the loosely bound electrons as free particles at rest, the estimated the cross-section can be written as ~\cite{Foot:2013msa,Foot:2002xz},
\begin{equation}
    \frac{d\sigma_{e^{\prime}}}{dE_R}=\frac{2\pi\epsilon^2\alpha^2 }{m_e E_R^2v_{min}^2}
\end{equation}
\begin{equation}
v_{min}=\sqrt{\frac{\left(m_e+m_{e^{\prime}}\right)^2E_R}{2m_em_e^{\prime^2}}}
\end{equation}
Where $\alpha$ is the fine structure constant. Assuming the mirror electron mass $(m_{e'})$ is identical to the Standard Model electron mass $(m_e)$, the minimum velocity required for the process in Eq.~\eqref{eq: 4.11} is $v_{min}=\sqrt{{2E_R}/{m_e}}$. The differential scattering rate is given by,
\begin{equation}
\frac{dR_{e^{\prime}}}{dE_R}=N_T N_f n_{e^{\prime}}\int_{v_{min}} \frac{d\sigma_{e^{\prime}}}{dE_R}f_{e'}(v)v d^{3}v
\end{equation}

Considering ${Y_{He'}}=0.74$ for kinetic mixing $\epsilon\sim 10^{-9}$, we estimate mirror electron number density in mirror plasma to be $n_{e'} \approx 0.2$ cm$^{-3}$ in the laboratory frame. Here $N_T$ denotes the number of target nuclei in detector. 
Setting,
\begin{equation}
\lambda=\frac{2\pi\epsilon^2\alpha^2 }{ E_R^2v_{min}^2}
\end{equation}
The differential interaction rate can be expressed as,
\begin{equation}
\frac{dR_{e^{\prime}}}{dE_R}= N_T N_f n_{e^{\prime}}\frac{\lambda}{E_R^2}\zeta_{e'}\left(v_{min}(E_R),t\right)
\end{equation}
$N_f$ represents the loosely bound electrons in target. These are defined as electrons with binding energies $E_B$ much smaller than the recoil energy $E_R$ $(i.e, E_B\ll E_R)$ allowing them to participate effectively in scattering. For germanium, we estimate 22 loosely bound electrons: 18 from the K shell and 4 from the N shell. The innermost M-shell electrons have $E_B\approx 180$ eV\footnote{  Binding energy data sourced from \url{https://xdb.lbl.gov/Section1/Table_1-1.pdf}} well below typical detector thresholds of (1-2) keV$_R$ and can thus be treated as free.  

For both mirror nuclei and mirror electrons, the recoil rates scale inversely with the squared recoil energy, a behavior distinct from WIMP scattering rates~\cite{Schumann:2019eaa,Peter:2013aha}. This difference arises because mirror DM particles interact electromagnetically via photon-mirror-photon kinetic mixing (governed by the strength parameter $\epsilon$). In such partially coherent scattering, sufficient nuclear recoil is expected to prevent excessive signal suppression, making mirror DM a viable candidate for direct detection.

\begin{figure}[htbp]
\centering
\includegraphics[width=.49\textwidth]{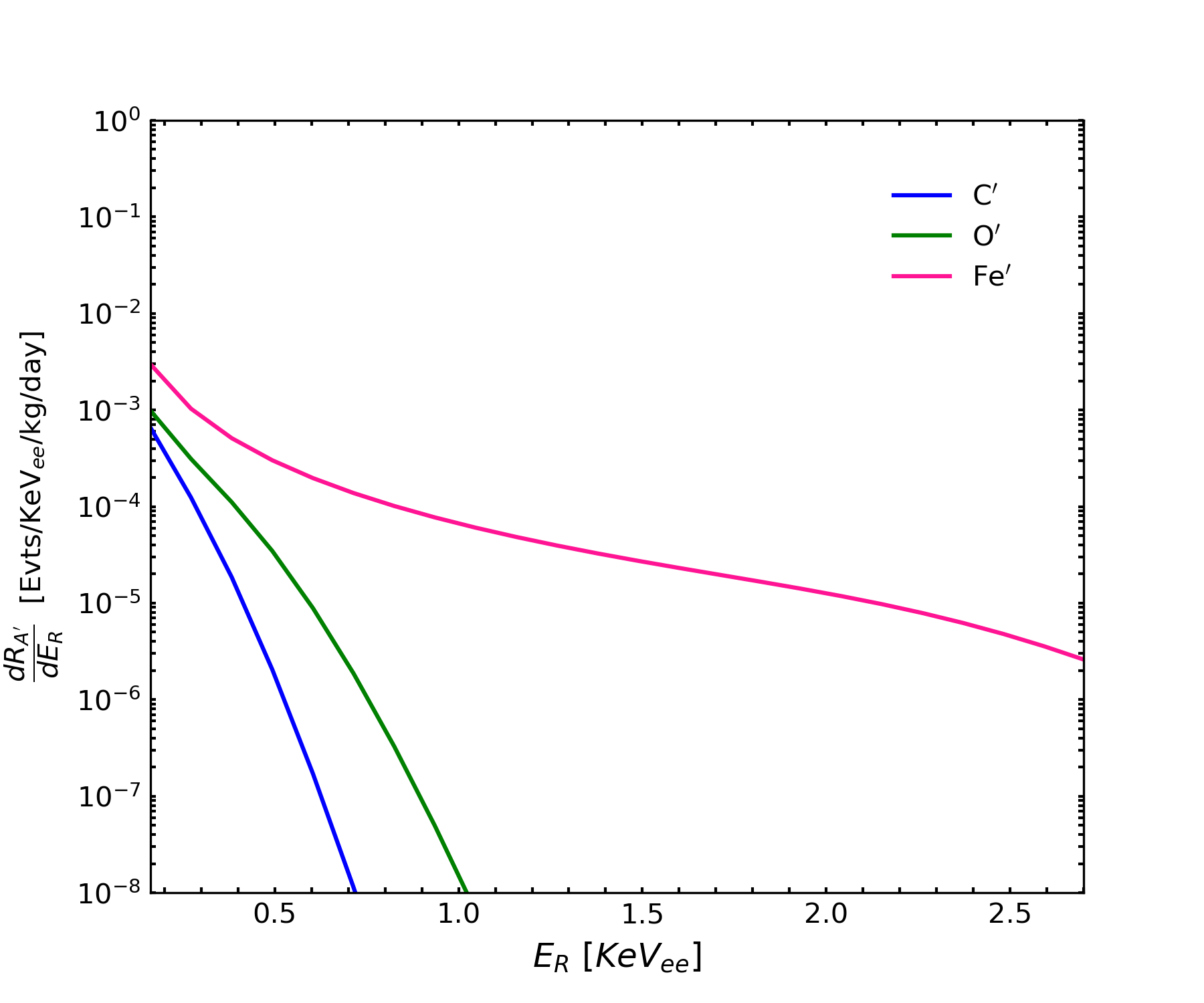}
\includegraphics[width=.49\textwidth]{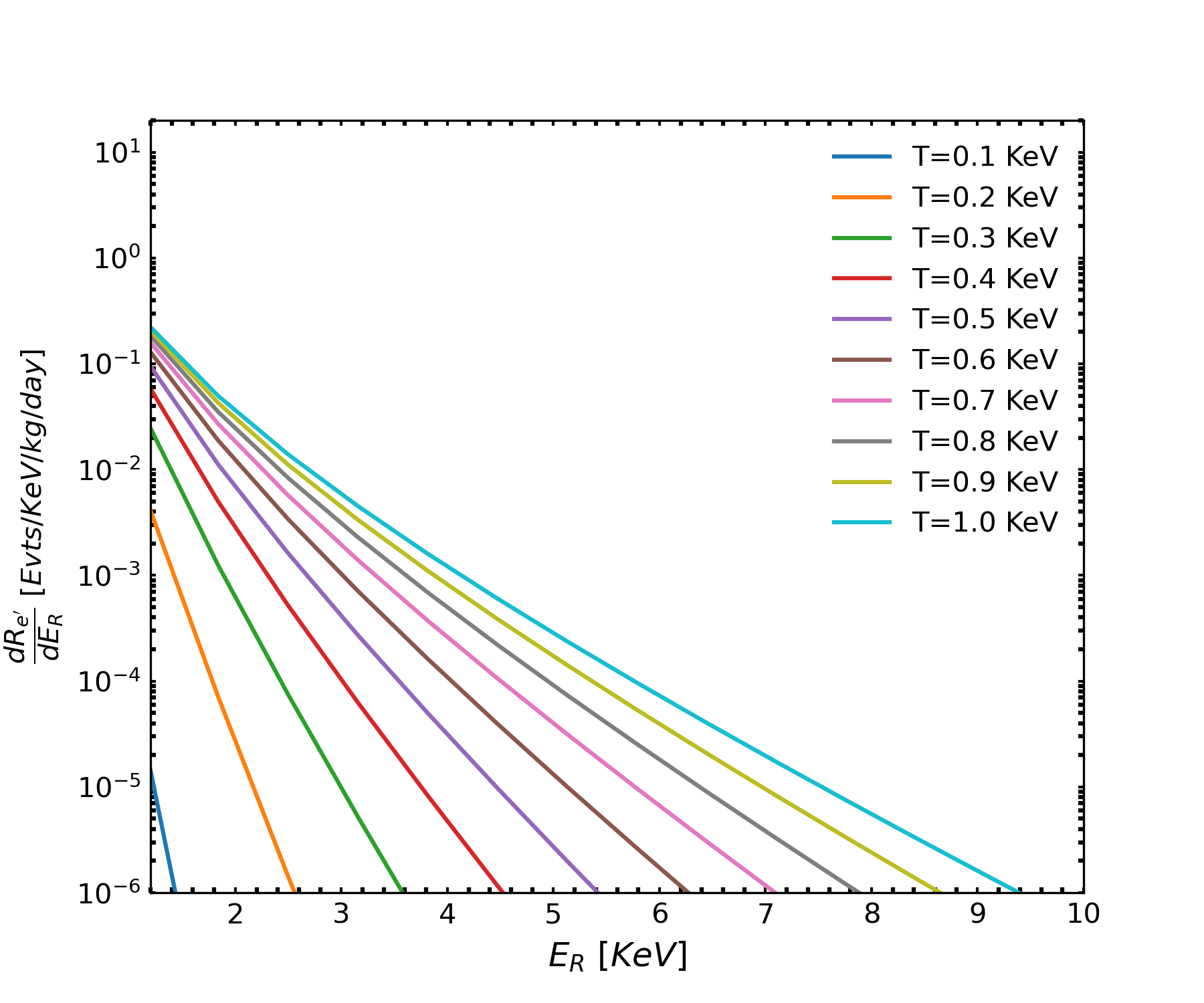}
\qquad
\centering
\caption{Left: Recoil rates from mirror nuclear DM with $\xi_{A^\prime} = 0.1\%$. Right: Mirror electron recoil rates for various halo temperatures $(T^{\prime})$. Both panels show CDEX-300 projections for 95 live days of exposure, assuming kinetic mixing $\epsilon = 5\times 10^{-11}$ and a Maxwellian velocity distribution with standard halo parameters ($v_{rot}, v_E) = (220, 239)$ km/s.} 
\label{fig:Fig2}
\end{figure}
\section{Experimental set up and limit evaluation}
\label{sec:Direct detection}
In their seminal work, Goodman and Witten proposed the concept of dark matter direct detection by searching for nuclear or electron recoils in detectors~\cite{Goodman:1984dc}. Mirror particles are expected to reach the detector and elastically scatter off target nuclei, producing nuclear or electronic recoils as discussed in Section~\ref{sec:Scattering formalism}. Since scattering involves the low momentum transfer, this require detectors with possibly a lower threshold and this is further justified by the fact that mirror DM has a smaller differential scattering cross-section at a higher threshold. In our phenomenological analysis, we incorporate the detector's characteristics i.e. resolution and efficiency, and smear the model prediction with the resolution of the detector which is customarily taken to be Gaussian. Thus the convoluted differential rate takes the form,

\begin{equation}
    \frac{d{R_c}}{dE_R}=\int \mathcal{R}(E_R|E_R') \frac{dR}{dE_R}(E_R^{\prime})dE_R^{\prime}
\end{equation}
where $\mathcal{R}(E_R|E_R')$ is the resolution function with energy dependent detector resolution $\sigma$,
\begin{equation}
\mathcal{R}(E_R|E_R')=\frac{1}{\sqrt{2\pi}\sigma(E_R^\prime)}e^{\left(-\frac{(E_R-E_R^\prime)^2}{2\sigma^2(E_R^{\prime})}\right)}
\end{equation}
 Using a bin-to-bin folding method, we reconstruct the theoretical recoil spectrum, summing individual bin contributions to obtain the total event rate in the recoil energy range. These binned reconstructions additionally account for the detector's energy-dependent efficiency. For direct detection experiments, where efficiency varies with recoil energy, the differential event rate takes the form,
\begin{equation}
    \frac{d{R_m}}{dE_R}=\kappa\,\frac{d{R_c}}{dE_R}
\end{equation}

 where $\kappa (E_R)$ is the efficiency of the detector. The total number of signal events in the energy range $E_R\in [E_R^{min}-E_R^{max}]$ is given by,
 \begin{equation}
     N=\int_{E_R^{min}}^{E_R^{max}} \chi\,\, \frac{d{R_m}}{dE_R} \,dE_R
 \end{equation}
 where $\chi$ denotes the effective exposure (typically in kg-year). 

\subsection{CDEX-10}
\label{CDEX-10}
Germanium detectors have been widely employed in rare event search experiments due to their excellent energy resolution and ionization yields. Germanium is a semiconductor, largely available in its pure form, and more crucial it can achieve the ultra-low radioactivity thresholds vital for rare-event detection~\cite{GERDA:2020emj, Majorana:2014jrt, SuperCDMS:2023sql, GERDA:2012qwd}.
   We access the data from the CDEX-10 analysis as reported in Refs.~\cite{CDEX:2022rxz,CDEX:2022kcd},
\begin{equation}
\sigma(E_{er})=a +\sqrt {b\cdot E}
\end{equation}
Here $\sigma(E_{er})$ is defined as the detector resolution and $a, b$ are constant factors read as $a = 0.35$ eV and $ b=2.8$ eV neglecting the small uncertainties in $a,b$. 
 Our analysis utilizes the recoil spectrum of $(0.16-4)$ keV$_{ee}$ with similar bin width to CDEX-10 analysis. The detector maintains high efficiency, reaching a combined efficiency of $4.5\%$ at the detection threshold 0.16 eV$_{ee}$~\cite{Jiang:2018lij, CDEX:2022fig}.

As we are interested in  photon-mirror-photon kinetic mixing for a given mirror DM mass and we implemented binned analysis. For each energy bin labeled by $k$, we estimate the number of expected events. Since rare event search experiments are typically limited by low statistics, it is evident that observed data follows a Poisson distribution. For example, if we denote the experimentally observed data points as $N^{obs}$ then the likelihood of $N^{obs}$ at given kinetic mixing $\epsilon$ and the signal events $N_k(\epsilon,m)$ can be expressed as,

\begin{equation}
\mathcal{L}( {N}^{obs}|\epsilon)=\prod_k \frac{N_k(\epsilon,m)^{N_k^{obs}}}{N_k^{obs}!} e^{-N_k(\epsilon,m)}
\label{eq:5.6}
\end{equation}

Where $\epsilon\neq 0$ means a physical process that corresponds presence of photon–mirror-photon kinetic mixing between SM and mirror sector and $\epsilon=0$ represents the absence of such mixing (and thus no signal). Since the $\epsilon=0$ case yields only background events, the likelihood for observing backgrounds can be written as,

 \begin{equation}
\mathcal{L}( {N}^{obs}|\epsilon=0)=\mathcal{L}_{bkg}
\label{eq: 5.7}
\end{equation}

We define the ratio of likelihoods from Eq. (\hspace{-0.1cm}~\ref{eq:5.6}) and Eq. (\hspace{-0.1cm}~\ref{eq: 5.7}) as,

\begin{equation}
    -2 \,\ln\left(\frac{\mathcal{L}( {N}^{obs}|\epsilon) }{\, \mathcal{L}_{bkg}}\right)= -2\sum_k\left[ N_k^{obs} \ln\left(\frac{N_k(\epsilon,m)+N_k^{bkg}}{N_k^{bkg}}\right)-N_k(\epsilon,m)\right]
\end{equation}

\begin{equation}
-2 \,\ln\left(\frac{\mathcal{L}( {N}^{obs}|\epsilon) }{\, \mathcal{L}_{bkg}}\right)=TS(\epsilon, m)=\chi^2
\end{equation}

 Here, TS denotes the test statistic which follows a $\chi^2$ distribution. In our analysis,
 the photon-mirror-photon kinetic mixing ($\epsilon$) is the only key parameter therefore we find $TS=\chi^2\approx 2.71$ at $90\%$ confidence level (C.L.) for one degree of freedom.
 
\subsection{CDEX-300}
\label{CDEX-300}
The forthcoming CDEX-300 experiment, scheduled for completion in 2027\footnote{\url{https://cdex.ep.tsinghua.edu.cn/}}, will employ a 300 kg enriched germanium detector array. With an anticipated active target mass reaching 1 ton within three years of operation, the experiment is designed to achieve ultra-low backgrounds ($\leq 5 \times 10^{-6}$ counts/kg/keV/year) and state-of-the-art energy resolution.

\begin{figure}[htbp]
\centering
\includegraphics[width=.49\textwidth]{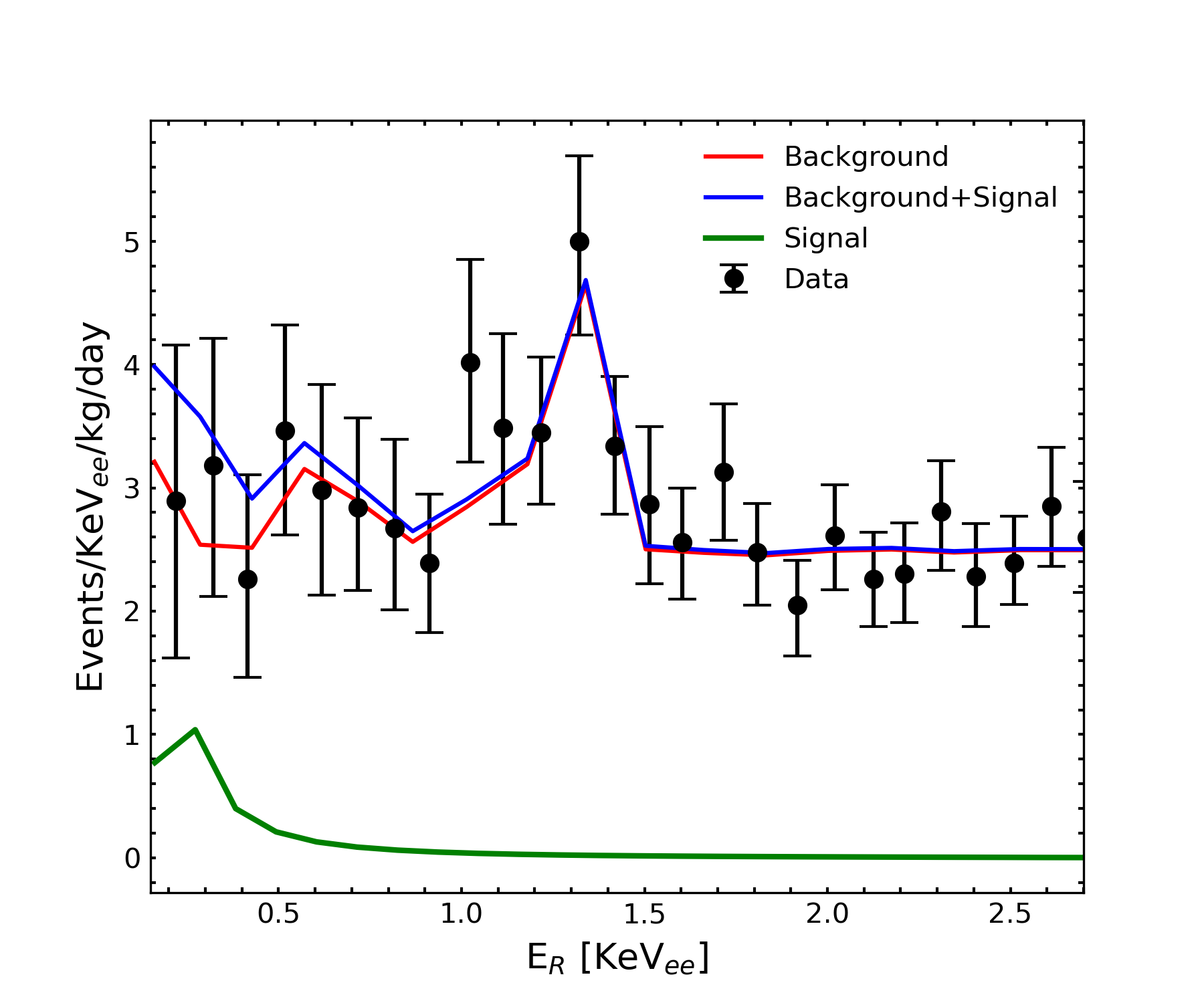}
\includegraphics[width=.49\textwidth]{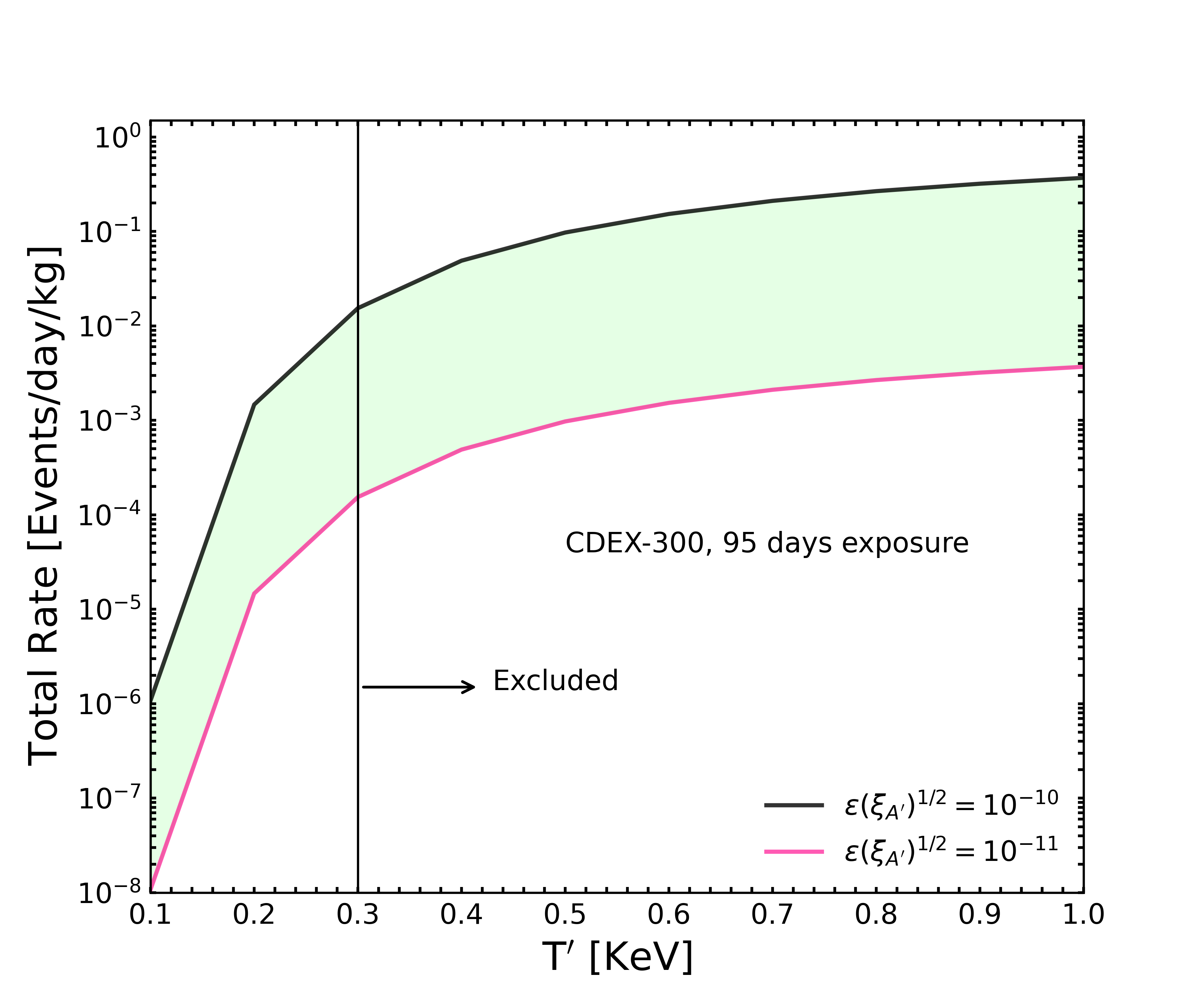}
\qquad
\centering
\caption{Left: The number of mirror nuclear DM events with respect to electronic recoil in CDEX-10, 205.4 days of exposure. Mirror iron DM is used as a signal model with kinetic mixing $\epsilon=2.45\times 10^{-10}$. Right: Total mirror electron rate in CDEX-300. The shaded band in lime color indicates theoretically motivated parameter space as discussed in Eq. (\hspace{-0.1cm}~\ref{eq: 2.3}). In both panel, we utilize Maxwellian VDF with halo parameters $(v_{rot}, v_{E}) = (220, 239)$ km/s.}
\label{fig:Fig3}
\end{figure}

We demonstrate CDEX-300's strong sensitivity to mirror DM through low-energy nuclear recoils (0.16–4 keV$_{ee}$). Our binned likelihood analysis incorporates optimized background rejection strategies, with future projections derived from Poisson statistics ($N=2.3$ at $90\%$ C.L.). Given its exceptional energy resolution and background suppression capabilities, CDEX-300 is poised to become a leading facility for probing kinetic mixing, as detailed in Sec. ~\ref{sec: CDEX-300 projection}.
\hspace{10cm}
\section{Results and discussion}
\label{sec:Results}
In this section, we present results for both the mirror electron and mirror nuclear DM as discussed in this work. Using the halo model with benchmark astrophysical and cosmological parameters, we calculate event rates while accounting for complementarity between CDEX-10 and CDEX-300. Our analysis focuses on probing the kinetic mixing parameter space through low-energy electronic recoils ($<5.1$ keV${ee}$), corresponding to 20 keV$_{R}$ nuclear recoils via Lindhard's model (see Appendix~\ref{app:Lindhard model}).
\subsection{Kinetic mixing limit from CDEX-10}
We present sensitivity projections for mirror nuclear DM interactions with  nuclei ranging from carbon to iron with atomic mass number ($A = 12-56$). The CDEX-10 detector demonstrates particular sensitivity to mirror nuclear dark matter (C$'$, O$'$, Fe$'$) due to reduced kinematic suppression from their larger masses, while substantially excluding lighter mirror nuclei (H$'$, He$'$). This implies that heavier mirror DM components even if subdominant in abundance can generate stronger detection signals, as quantified in Table~\ref{tab: Tab.1}.
\begin{figure}[htbp]
\centering
\includegraphics[width=.47\textwidth]{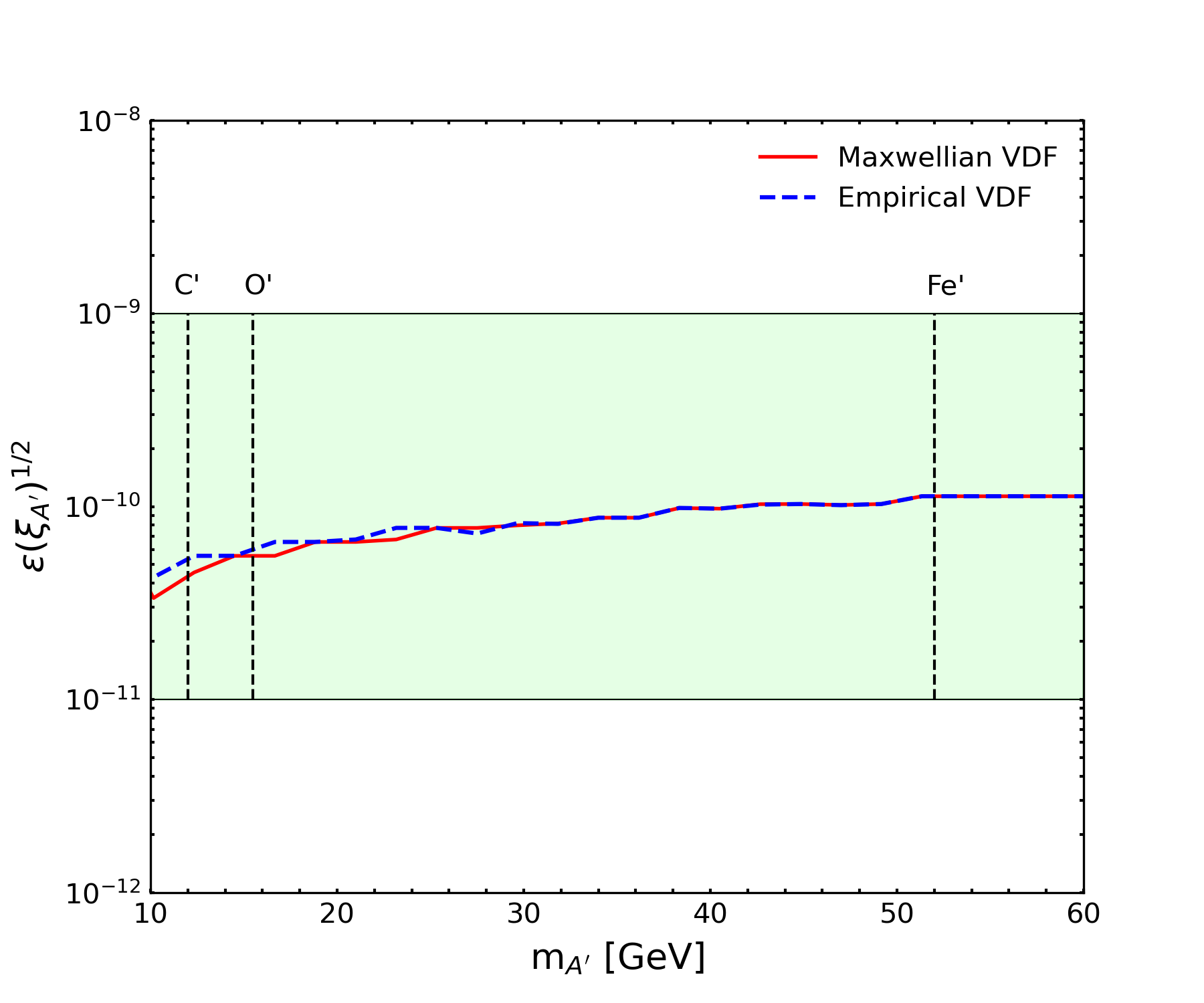}
\includegraphics[width=.47\textwidth]{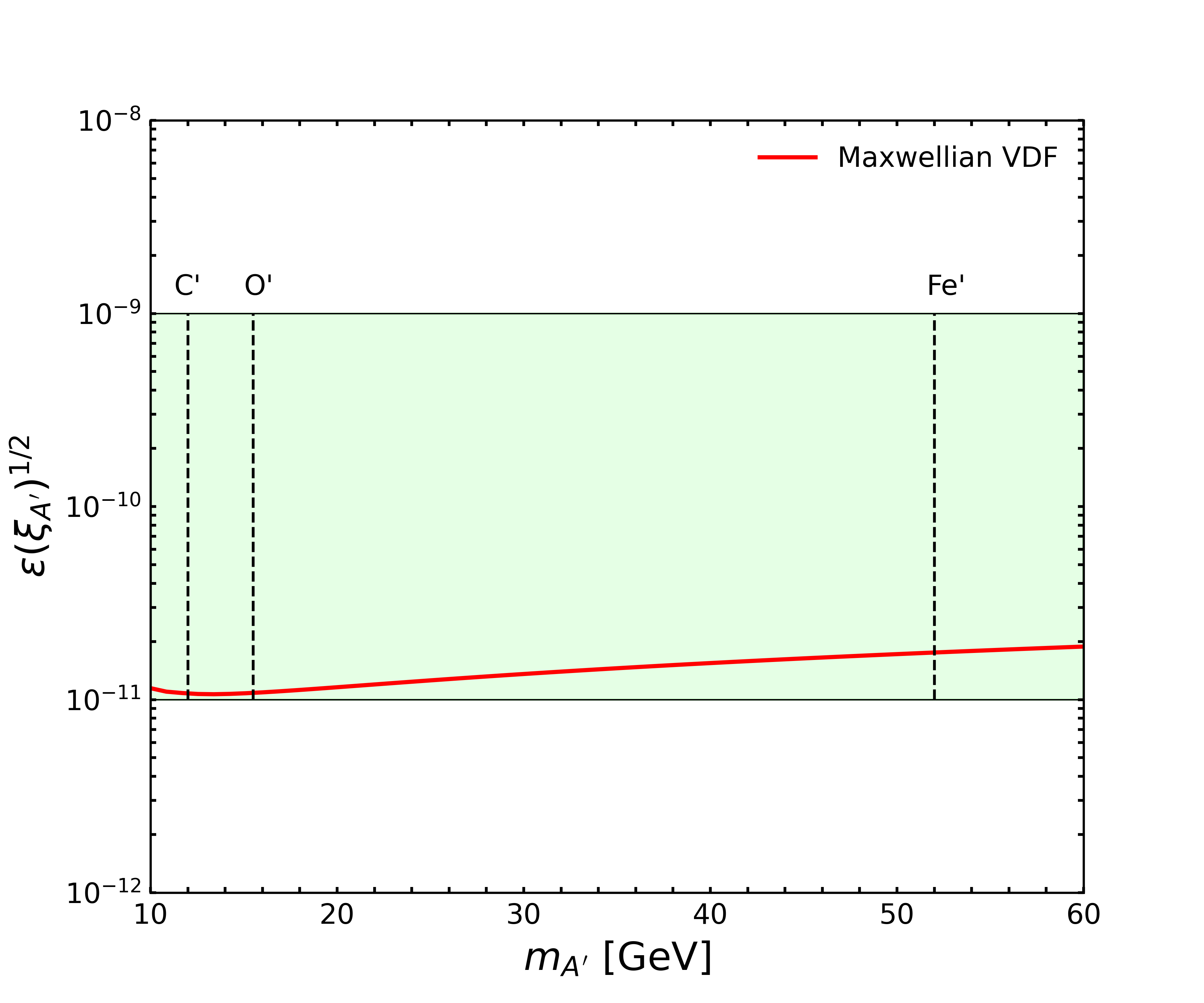}
\qquad
\centering
\caption{ Left: $90\%$ C.L. exclusion limits on photon-mirror-photon kinetic mixing ($\epsilon$) from 205.4 days of CDEX-10 data, comparing Maxwellian (solid red) and empirical (blue dashed) dark matter velocity distributions. Right: Projected $90\%$ C.L. sensitivity for mirror nuclear DM, with the shaded band indicating theoretically favored parameter space as discussed in Eq. (\hspace{-0.1cm}~\ref{eq: 2.3}). Both analyses assume standard halo parameters $(v_{rot}, v_{E}) = (220, 239)$ km/s. } 
\label{fig:Fig4}
\end{figure}

It can be seen from Fig. \ref{fig:Fig4} that CDEX-10  data in the low mass region i.e $(12- 20)$ GeV exhibit a comparatively
better reach to the kinetic mixing than the $(20-56)$ GeV region for both Maxwellian and empirical velocity distribution functions (VDFs). This can be understood by the fact that at higher recoil, the tail of the mirror velocity distribution is strongly suppressed as a consequence less nuclear recoil is anticipated therefore sensitivity decreases regardless of detector threshold, energy resolution, efficiency, and halo model parameters. In our analysis, we have taken into account galactic rotational velocity $v_{rot}\in(190,250)$ km/s with an average value of Earth velocity $v_E\equiv 239$ km/s as data has been taken from February to August within a year of CDEX-10 run. We observed that the model results in almost the same number of events in a variation of $v_{rot}$ and these behaviors are justified by
the fact that velocity distribution is overall less sensitive to changes in galactic halo parameters which cause up to less than $2\%$\footnote{Percentage change in kinetic mixing sensitivity is calculated in respect of {$(v_E, v_{rot})=(239,220)$ km/s with the choices of sets $(v_E, v_{rot})=(239,250)$ km/s and $(v_E,v_{rot})=(239,190)$ km/s}} effect in kinetic mixing in the allowed region of parameter space.  As the choice of halo model parameters has a sub-leading effect on kinetic mixing and is washed out at higher recoil therefore the exclusions limit denoted by red and dotted blue lines corresponding to two different mirror velocity distributions as shown in the left panel of Fig. \ref{fig:Fig4} lies on each other is meaningful and expected. With careful investigation of the current CDEX-10 data, we find the kinetic mixing $\epsilon>2\times 10^{-10}$ excluded irrespective of the choice of halo model parameters and mirror nuclei i.e C$'$, O$'$, Fe$'$ in the signal model while $\epsilon\leq ( 10^{-11}-2\times 10^{-10})$
remain viable in the mass range of $(10-56)$ GeV.  In addition, we examine the choice of mirror form factors. The measure of nuclear radius and nuclear skin thickness associated with mirror dark matter, we observed these choices can cause less than $2\%$ differences in signal events. A $90 \%$ exclusion limit on the photon-mirror-photon kinetic mixing calculated using Lewin-Smith form factors generated within the Helm parameterization matches the Engel form factors parameterization and the ratio of these estimated sensitivities is close to one. Therefore, we suggest that either of the choices of parameterizations in mirror nuclear form factors will suffice in determining the kinetic mixing. We project $90\%$ upper limits on kinetic mixing of $\epsilon < 3 \times 10^{-11}$ for mirror nuclear DM which is an order of better agreement than the derived limit from CDEX-10 in this work. It is noted that we do not perform mirror electron analysis with CDEX-10 data because the mirror electron model predicts a large number of events and given CDEX-10 data the events statistics are low therefore better constraints can not be achieved however this analysis can be performed with CDEX-300 due to its 30 times bigger fiducial volume than CDEX-10. In the following section, we present sensitivity projections for the CDEX-300 with mirror electron model,  exploring how the mirror sector's temperature can be constrained through photon-mirror-photon kinetic mixing.

\subsection{Sensitivity projection at CDEX-300}
\label{sec: CDEX-300 projection}
  The mirror electron model is fully characterized by two key parameters: the kinetic mixing between photon-mirror-photon denoted by parameter $\epsilon$ and the local temperature $T'$, which quantifies the thermal energy distribution of mirror particles in mirror halo. Previous experimental constraints on mirror sector parameters include:
(i) temperature limits from LUX, reporting $T'\sim 0.3$ keV for kinetic mixing $\epsilon\sim 10^{-11}$~\cite{LUX:2019gwa}, and
(ii) kinetic mixing bounds ($\epsilon < 5.8 \times 10^{-8}$) from positronium decays to invisible states in vacuum~\cite{Vigo:2019bou}.
\begin{figure}[htbp]
\centering
\includegraphics[width=.65\textwidth]{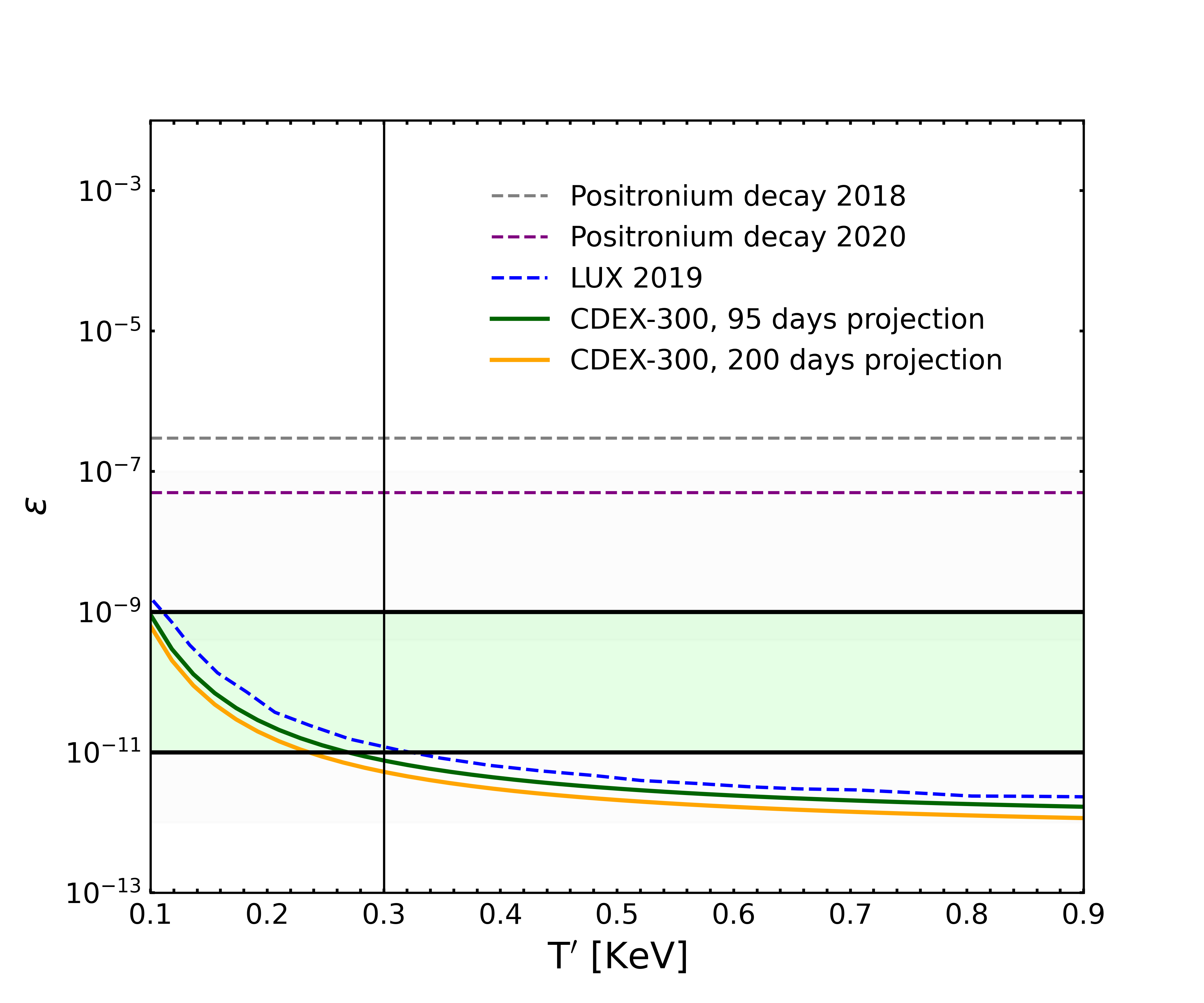}
\caption{ $90\%$ C.L. projected sensitivity on the kinetic mixing parameter $\epsilon$ for mirror electron DM assuming $\epsilon=5\times 10^{-11}$ and Maxwellian VDF with halo parameters $(v_{rot}, v_{E}) = (220, 239)$ km/s. The shaded region indicates theoretically allowed parameter space as discussed in Eq. (\hspace{-0.1cm}~\ref{eq: 2.3}). Dashed grey and purple curves show existing constraints from positronium decay experiments~\cite{Vigo:2018xzc, Vigo:2019bou}.}
\label{fig:Fig5}
\end{figure} 
Our findings suggest CDEX-300 with an exposure of 95-200 days  with $160$
eV$_{ee}$ threshold can significantly constrain the parameter space below the local temperature of 0.3 keV and above the local temperature of 0.3 keV, kinetic mixing can be explicitly excluded. As the Helium I and Helium II ionization energies in their ground state $  1s^2\; ^1\!S_0$ and $  1s^2\; ^1\!S_{1/2}$  are around 24.6 eV and  54.42 eV, it is anticipated that at the temperature around 0.28 keV, the halo composed of mirror particles with dominant abundance in helium can be completely ionized. This further motivates the presence of the mirror electrons in the halo and the scattering we discussed in Sec.~\ref{Mirror electron scattering} is reasonable to happen in the detector. Certainly, the available local temperature in the mirror halo is directly scaled with the mean mass of mirror particles and their velocity dispersions, a non-zero value of helium fraction advises non-zero photon-mirror-photon kinetic mixing between ordinary and mirror electromagnetism $U(1)_Y$ and $U(1)^\prime_Y$. Current theoretical frameworks suggest the mirror sector is cooler than the ordinary sector. In fact, the BBN bound on the effective number of extra
light neutrinos imply that the mirror sector has a temperature lower than the ordinary sector and a ratio of temperatures between these two sectors is often described by a free parameter x as discussed in Ref.~\cite{Ciarcelluti:2009da},
\begin{equation}
    x=\frac{T'}{T}\approx 0.2\left(\frac{\epsilon}{10^{-9}}\right)^{1/2}
    \label{eq:6.1}
\end{equation}

Here $T\leq 1$ MeV is the temperature in the ordinary sector and $T'$ is the temperature in the mirror sector.
Cosmological constraints from CMB measurements and large-scale structure observations indicate that a temperature ratio between the mirror and standard sectors is $\frac{T'}{T }<0.3$ with a value of kinetic mixing $\epsilon<3\times 10^{-9}$ which is further studied by DAMA in explanation of annual modulation~\cite{Foot:2008nw, Ciarcelluti:2004ip, Berezhiani:2003wj, Berezhiani:2003xm, Berezhiani:2008zza, Addazi:2015cua}. We project upper limits on kinetic mixing of $\epsilon \leq 3.75 \times 10^{-11}$ for mirror electrons at a mirror sector temperature $T' \leq 0.28$ keV. These new constraints significantly improve upon existing limits from both LUX and positronium decay experiments, as demonstrated in Fig.~\ref{fig:Fig5}.  Since mirror DM requires energy input from ordinary supernovae, theoretical studies suggest that kinetic mixing $\epsilon \sim 10^{-9}$ is necessary for energy supply to the mirror sector~\cite{Foot:2016wvj}. 
 Our projected sensitivity of $\epsilon \sim 10^{-11}$ implies that kinetic mixing at this level would be too weak to efficiently transfer energy to the mirror sector, potentially disrupting the spherical halo in spiral galaxies.
\section{Conclusion}
\label{sec:Conclusion}
In this work, we study the theory of mirror DM halos composed of collisionless mirror particles and analyze their local velocity distribution, beginning with a Maxwellian approximation and then introducing an empirical model tailored for mirror-sector kinetic mixing. Incorporating nuclear form factors (for both mirror and germanium nuclei), halo model parameters, and associated astrophysical uncertainties, we derive new constraints on photon-mirror-photon kinetic mixing using CDEX-10 low-energy recoil data. Furthermore, we present sensitivity projections for CDEX-300, demonstrating its capability to probe theoretically motivated regions of the mirror DM parameter space. With recoil energy thresholds comparable to CDEX-10, a larger detector volume, and significantly reduced backgrounds, CDEX-300 will place stringent constraints on kinetic mixing.
\appendix
\section{Lindhard's model}
\label{app:Lindhard model}
In dark matter direct detection, often we require to calculate electronic equivalent energy $(E_{ee})$ and corresponding nuclear recoil $(E_{R})$ in the detector. One can convert nuclear recoil to electronic equivalent and vice versa using the Lindhard model as follows~\cite{osti_4701226,lindhard},

\begin{equation}
L(E_{R})=\frac{k\,g(\eta)}{1+k \,g(\eta)}
\end{equation}
\begin{equation}
   g(\eta)=3\eta^{1.5}+0.7\eta^{0.6}+\eta 
\end{equation}
\begin{equation}
E_{ee}=L(E_{R})\cdot E_{R}
\end{equation}
with $\eta=11.5$ (E$_{R}$/KeV)Z$^{-7/3}$. The quantity $g(\eta)$ is the function of the recoil can be computed using the Thomas-Fermi screening
function. 
The parameter k is a proportionality constant and the optimized value of $k = 0.177\pm 0.0060$ is reported in ~\cite{Sorensen:2011bd}. Typically, it falls within the range $(0.1,0.2)$.

\acknowledgments
I am grateful to Sunny Vagnozzi for insightful discussions on this work. I sincerely thank Vedran Brdar for his valuable suggestions during manuscript preparation. I also acknowledge Michael Gold, and Micheal Graesser for helpful correspondence.



 \bibliographystyle{JHEP}
\bibliography{biblio.bib}






\end{document}